\documentstyle[12pt,epsfig]{article}
\def \ins#1#2#3#4#5#6 {
  \begin{figure}[#1]
    \begin{center}
      \psfig{file=#2,width=#3,height=#4,angle=0}
      \caption{#5}
      \label{#6}
    \end{center}
  \end{figure}
  }

\begin{document}
\large

\begin{center}
{\Large \bf
Web-page on UrQMD Model Validation
}
\end{center}

\begin{center}
A. Galoyan,
J. Ritman\footnote{University of Bochum and Research Center Julich,
                   Institute for Nuclear Physics (IKP),
                   Forschungszentrum Julich, Germany},
V. Uzhinsky\\
Joint Institute for Nuclear Research,\\
Dubna, Russia
\end{center}

\begin{center}
\begin{minipage}{12cm}
A WEB-page containing materials of comparing experimental data and
UrQMD model calculations has been designed. The page provides its user with a variety of tasks
solved with the help of the model, accuracy and/or quality of experimental data description,
and so on. The page can be useful for new experimental data analysis, or new experimental research
planning.

\ \ \ \ The UrQMD model is cited in more than 272 publications. Only 44 of them present original
calculations. Their main results on the model are presented on the page.

\end{minipage}
\end{center}

\vspace{0.5cm}

\newpage

Monte Carlo event generators play a very important role in high energy physics. One can mark
the following areas of their application:

\begin{enumerate}

\item {\bf Pragmatics or practical tasks} -- development of new or upgrade of old experimental
setups to study some processes/interactions, design of detectors, Monte Carlo simulation of
the detector responses and so on.  Event generators applied at these should have fast
operation speed, stability of work, and a rough reproduction of previous experimental results.
As an example, let us mention applications of the UrQMD model \cite{UrQMD1,UrQMD2} for the
development of detectors for research on nucleus-nucleus interactions
(CBM collaboration \cite{CMB-home-page}), and detectors for investigation of antiproton-proton
annihilations (PANDA collaboration \cite{PANDA-home-page}) at future GSI accelerators. The
RQMD and HIJING models have been used for analogous purposes for RHIC experiments. The
well known GEANT package \cite{GEANT} is widely used to simulate various installations.

\item {\bf Analysis of new experimental data and new investigations planning}. They include
comparison of new data with previous data and model predictions. As a rule, the new data do
not agree with model predictions, so  some questions arise in this regard: whether
all special features of the setup have been taken into account; whether they are free of
methodical errors; whether the theoretical models are used correctly; whether the model
parameters set is right; whether the discrepancy between the experimental data
and model predictions is of systematical character; whether the discrepancy has been observed
in previous experiments; whether the discrepancy has been considered as an evidence of a
new physical effect, and so on. They are solved differently, and often quite difficult using
quite often event generators. The generators should have a flexibility in parameters variation and
physical scenario, as well as a sufficient physical meaning of the parameters.

~~~~~Another situation takes place at new research planning. A first question asked by
experimentalists is related to the load of the setup by ordinary background processes.
A second deals with the radiation condition of the experiment. A third question asks about
the admixture of the background processes in the phenomenon under study, and how it can be
damped, and so on. Clearly, experimentalists prefer to use well approbated and well recommended
models for their  estimations. Here one can not avoid a study of model application
experience. As a rule, there is no enough time for this. Thus, experimental collaborations
attract the authors of the models or use authors variants of model code to solve
the questions. As an example, let us point out on estimations of secondary particles multiplicity
in central gold-gold interactions obtained by CBM collaboration \cite{CBM_TPR}, and estimations
of background processes intensities of the PANDA collaboration \cite{PANDA_TPR}.

\item {\bf Scientific or cognition aims only} -- search for new effects or phenomena on
the base of analysis of a discrepancy between experimental data and model predictions. One uses
the fact that Monte Carlo models are a synthesis of existing notions about process mechanics.
Thus, the discovered discrepancy can be considered as an evidence of our insufficient understanding,
or as an evidence of new effects. For example, the discrepancy between experimental data and
intra-nuclear cascade model calculations growing with the collision energy rising led in its time
to appearing a very important conception for high energy physics -- "formation time of
secondary particle".

\end{enumerate}

The final aim of all the efforts is creation of a theory of processes that could predict
effects with any predetermined exactness. As there are only few such theories, the
aim is re-formulated -- creation of a theory or a model predicting observable effects with
specified exactness. Determination of the exactness is a special additional task.

The philosophical aspect of the scientific research -- "cognition of Good wisdom", is out of
the scope of our consideration.

The aim of this work is to create a WEB-page containing materials for the second trend
on the well-known Ultra-relativistic Quantum Molecular Dynamics model\footnote{
A short description of the model see in Appendix 1}$^)$ application. This foresees the
following: to compose a list of papers where the model is mentioned, to select papers with
original results on the model, to classify materials and to create their graphical
representation.

According to the electronic database of scientific publications\\
(http://www-spires.fnal.gov/spires/hep/search/), the milestone papers on the UrQMD model
\cite{UrQMD1,UrQMD2} were mentioned in 272 publications at the beginning of year 2006. All of them
were looked through, and some of them presenting materials on the model were selected.
The last publications were studied for original calculations with
model usage. Only 44 publications were selected (the list of the papers is given in
Appendix 2). Graphical materials from the papers were put on the WEB-page.

All published calculations in comparison with experimental data were sorted according to the
following sections.

\begin{enumerate}

\item Production cross sections

\item Particle multiplicities

\item Particle multiplicity ratios

\item Multiplicity distributions and correlations

\item Rapidity distributions of pions, kaons, protons and others

\item $m_T$-distributions, temperature

\item $P_T$-distributions

\item Flow

\item Event-by-Event fluctuations

\item Dileptons, J/Psi production

\item Others

\end{enumerate}

Three subdivisions were introduced in each of the section -- hadron-nucleon, hadron-nucleus,
and nucleus-nucleus interactions. As a result, the main page looks as it is presented in Fig.~1.
\ins{cbth}{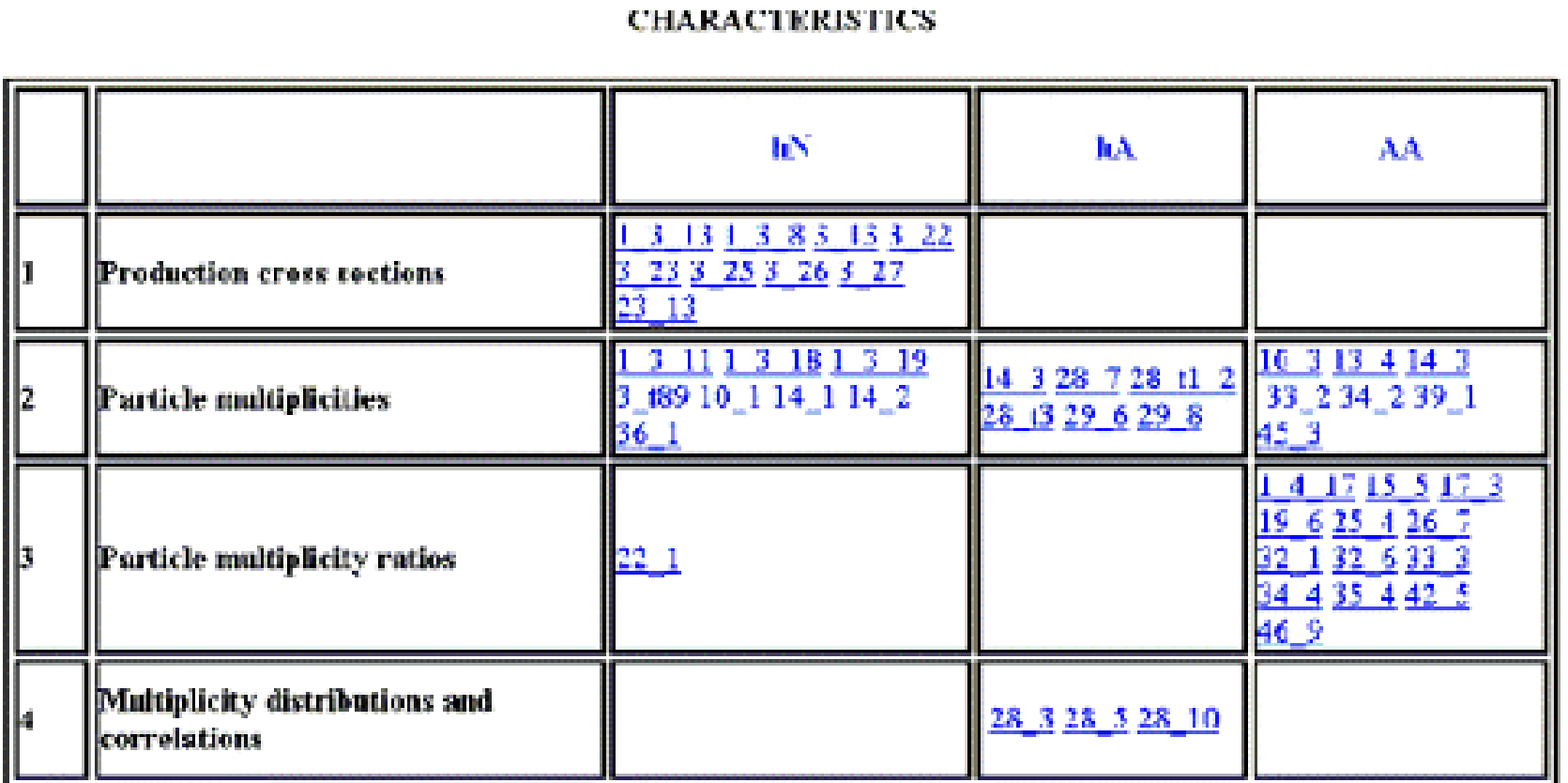}{160mm}{80mm}{View of the main page.}{fig1} 

Each cell of the table contains references on the pages with corresponding materials.
Each of the reference has a form XX\_YY, where XX is an order number of the paper in the list of
used papers, and YY is an order number of the figure in the paper.

Section 1.1 (Production cross sections, hh-interactions) collects descriptions of cross
sections of various reactions in hadron-nucleon collisions. Mainly, they have been presented
in the first publications \cite{UrQMD1,UrQMD2}. Note that the set of the cross sections represents
only a small part of cross sections collected in the well-known compilation \cite{Flaminio}.
One can think that an analysis of a larger set of experimental data allows one to define more
precisely the boundary between the binary model of the hadron-hadron interactions implemented
in the UrQMD model at low energies and the FRITIOF model \cite{FRITIOF} used at high energies.
This may improve the ratio between the cross sections of multi-meson and few-meson reactions
in the UrQMD model.

Section 2 (Particle multiplicities) gives calculations of multiplicities of $\pi$-mesons,
$K$-mesons, protons, anti-protons and other particles. It is on the alert that
the model underestimates the multiplicities of $\pi$-mesons, and overestimates the multiplicities
of strange baryons, anti-protons in NN-interactions (see pages 1\_3\_11, 3\_t89, 14\_2). At
the same time, the model overestimates meson multiplicities and underestimates mutiplicities
of protons, anti-protons, and hyperons in nucleus-nucleus collisions (see Section 2.3).

In Section 3 (Particle multiplicity ratios) relative multiplicities of
$K^+/\pi^+$, $K^-/\pi^-$ etc. are considered. They get actual due to the papers
\cite{Gazd1,Gazd2,Gazd3}. The problem with description of $K^+/\pi^+$ ratio in nucleus-nucleus
interactions (see page 34\_4) at energies larger than 10 GeV allows one to assume that
at high energies an additional transverse pressure appears in nucleus-nucleus interactions
\cite{34} which is not implemented in the UrQMD 1.3 model.

In Section 4 (Multiplicity distributions and correlations) calculations of particle multiplicity
distributions and correlation dependencies between particle multiplicities are collected. They
can be of interest for experiments at sufficiently low energies with a good particle identification.
Particle multiplicity dependence on the impact parameter in nucleus-nucleus interactions,
correlations between multiplicities of particles produced in different phase-space regions,
correlations between energies of neutral and charged particles etc. can be put in the section.
Corresponding calculations are absent but the properties can be measured or have been measured
in experiment.

The main volume of results is presented in the Sec. 5 (Rapidity distributions). Here $K^+$-meson
rapidity distribution in $pp$ interactions at 160 GeV is given which is in a bad agreement with
experimental data (see page 23\_2). At the same time, analogous theoretic calculations for
nucleus-nucleus interactions agree with corresponding experimental data. Baryon rapidity
distributions in $pp$-interactions as well as in nucleus-nucleus collisions at energies larger
than 6 GeV/nucleon are badly described by the model, especially in the nuclear fragmentation
regions (see 13\_3, 23\_1 and 24\_1). The worst situation takes place with the description of
anti-proton production in $PbPb$-interactions at 160 GeV/nucleon.

Section 6 ($m_T$-distributions, temperature) deals with particle distributions on transverse
masses, $m_T$, and related estimations of particle source temperatures. According to the collected
materials, meson spectra in $pp$ and $pA$ interactions are described quite well by the model.
The problem exists with $K$-meson spectrum reproduction for nucleus-nucleus
interactions at energies larger than 6 GeV/nucleon. One can think that one needs to take into
account the creation of quark-gluon-plasma for its solution.

In Section 7 ($P_T$-distributions) not numerous calculations of properties of particle
transverse momentum distributions are collected. The calculations of photon transverse momentum
distributions in $PbPb$-interactions are also presented here (see page 21\_10). Since the existing
version of the model (1.3) does not consider the so-called "hard" processes, one can not expect
a good description of the distributions at $P_T\geq 2$ in the nearest future.

Section 8 (Flow) presents calculations of characteristics of flow in nucleus-nucleus interactions
($v_1, \ v_2,\ p_x,\ F$). Taking into account that this trend gets quite actual after the RHIC
result publication, the volume of corresponding model calculations does not look sufficient. In
particular, there is no analysis of $v_2$ dependence on $P_T$ or centrality of interactions. In
order to recover this deficiency, we present calculations of $v_2(P_T)$ (pages 44\_6 and 44\_7)
without comparison with corresponding experimental data.

Section 9 is devoted to Event-by-Event fluctuations. The characteristics are expected to
be actual in the nearest future.

Section 10 collects characteristics of dileptons produced in hardon-nucleus and
nucleus-nucleus interactions.

Finally, various seldom calculations are presented in Sec. 11.

The page is available at\\
http://hepweb.jinr.ru/urqmd1\_3/validation/urqmd\_model\_validation.htm

The authors are thankful to Prof. G.A. Ososkov and Prof. Kh. Abdel-Waged for a reading of
the paper and important remarks.

\section*{Appendix 1: Short description of the UrQMD model}

The Ultra-relativistic Quantum Molecular Dynamic model (UrQMD)
\cite{UrQMD1,UrQMD2} is a microscopic model based on a
phase space description of nuclear reactions. It describes the
phenomenology of hadronic interactions at low and intermediate
energies ($\sqrt s <5$~GeV) in terms of interactions between known
hadrons and their resonances. At higher energies, $\sqrt s >5$~GeV,
the excitation of color strings and their subsequent fragmentation
into hadrons are taking into account in the UrQMD model. The model
was proposed mainly for a description of nucleus-nucleus interactions.
Note, that up to now there is no unique theoretical description of
the underlying  hadron$-$hadron interactions, with their vastly
different characteristics at different incident energies and in
different kinematic intervals. Perturbative quantum chromodynamics
(pQCD) can be applied to describe hard processes, i.e. processes
with large four-momentum, $Q^2$, transfer. But pQCD is formally
inappropriate for the description of the soft interactions because of
the absence of the large $Q^2 -$scale. Therefore, low$-p_T$ collisions
are described in terms of phenomenological models. A vast variety of
models for hadronic- and nuclear collisions have been developed. The
UrQMD model is the most appropriate one for energy range of the future
PANDA and CBM  experiments.

The model is based on the covariant propagation of all hadrons
considered on the (quasi-)particle level on classical trajectories in
combination with stochastic binary scatterings, color string
formation and resonance decay. It represents a Monte Carlo solution of a
large set of coupled partial integro-differential equations for the time
evolution of the various phase space densities of particle species
$i=N,\Delta,\Lambda,$ etc. The main ingredients of the model
are the cross sections of binary reactions, the two-body potentials and
decay widths of resonances.

In the model, as in other Quantum Molecular Dynamics (QMD) models
(see \cite{aichelin91a,aichelin87b,aichelin86a,peilert89a}),
each nucleon is represented by a coherent state of the form ($\hbar,c =1$)
\begin{equation}
\label{gaussians}
\phi_i (\vec{x}; \vec{q}_i,\vec{p}_i,t) =
\left({\frac{2 }{L\pi}}\right)^{3/4}\, \exp \left\{
-\frac{2}{L}(\vec{x}-\vec{q}_i(t))^2 + \frac{1}{\hbar} {\rm i} \vec{p}_i(t)
\vec{x} \right\}
\end{equation}
which is characterized by 6 time-dependent parameters,
$\vec{q}_{i}$ and $\vec{p}_{i}$, respectively.
The parameter $L$, which is related  to the extension of the wave packet in
phase space, is fixed.
The total $n$-body wave function is  assumed to be the direct
product of coherent states (\ref{gaussians})
\begin{equation}
\Phi = \prod_i \phi_i (\vec{x}, \vec{q_i}, \vec{p_i}, t)
\end{equation}

The equations of motion of the many-body system is calculated by means
of a generalized variational principle. The Hamiltonian $H$ of the system
contains a kinetic term and mutual interactions $V_{ij}$
($H = \sum_i T_i + {\ \frac{1 }{2}} \sum_{ij} V_{ij}$). The time
evolution of the parameters is obtained by the requirement that the action
is stationary under the allowed variation of the wave function.
This yields an Euler-Lagrange equation for each parameter.
\begin{equation}\label{hamiltoneq}
\dot{\vec{p}}_i = - \frac{\partial \langle H \rangle}{\partial \vec{q}_i}
\quad {\rm and} \quad
\dot{\vec{q}}_i = \frac{\partial \langle H \rangle}{\partial \vec{p}_i} \, .
\end{equation}
\begin{equation}
\dot{\vec{q}_i} = {\frac{\vec{p}_i }{m}} + \nabla_{\vec{p}_i} \sum_j
\langle V_{ij}\rangle  = \nabla_{\vec{p}_i} \langle H \rangle
\end{equation}
\begin{equation}
\dot{\vec{p}_i} = - \nabla_{\vec{q}_i} \sum _{j\neq i} \langle
V_{ij}\rangle = -\nabla_{\vec{q}_i} \langle H \rangle
\end{equation}
\begin{equation}
\langle V_{ij}\rangle = \int d^3x_1\,d^3x_2\,
\phi_i^* \phi_j^* V(x_1,x_2) \phi_i \phi_j
\end{equation}
These are the time evolution equations which are solved numerically.
The equations have the same structure as the classical Hamilton equations.

The interaction is based on a non-relativistic density-dependent
Skyrme-type equation of state with additional Yukawa- and Coulomb potentials.
Momentum dependent potentials are not used -- a Pauli-potential, however,
may be included optionally. The Skyrme potential consists of a
sum of two- and a three-body interaction terms. The two-body
term, which has a linear density-dependence models the long range attractive
component of the nucleon-nucleon interaction, whereas the three-body
term with its quadratic density-dependence is responsible for the
short range repulsive part of the interaction. The parameters of the
components are connected with the nuclear equation of state. Only the
hard equation of state has been implemented into the current
UrQMD model.

A projectile or target nucleus is modeled according to the Fermi-gas ansatz.
The wave-function of the nucleus is defined as the product wave-function
of the single nucleon Gaussians. In configuration space the centroids
of the Gaussians are randomly distributed within a sphere with radius $R(A)$,
\begin{equation}
R(A) \,=\, r_0 \left( \frac{1}{2} \left[ A + \left( A^{\frac{1}{3}} -1
        \right)^3 \right] \right)^{\frac{1}{3}} \, ~~~~~
r_0 \,=\, \left( \frac{3}{4 \pi \rho_0} \right)^{\frac{1}{3}} \, .
\end{equation}
$\rho_0$ is the nuclear matter ground state density used in the UrQMD model.

The phase-space density at the location of each nucleon is evaluated
after its placement. If the phase-space density is too high (i.e. the
respective area of the nucleus is already occupied by other nucleons),
then the location of that nucleon is rejected and a new location is
randomly chosen.

The initial momenta of the nucleons are randomly chosen between 0 and
the local Thomas-Fermi-momentum:
\begin{equation}
p_F^{max} \,=\, \hbar c
        \left( 3 \pi^2 \rho \right)^{\frac{1}{3}} \, ,
\end{equation}
with $\rho$ being the corresponding local proton- or neutron-density.

A disadvantage of this type of initialization is that the initialized
nuclei are not really in their ground-state with respect to the
Hamiltonian used for their propagation. The parameters of the Hamiltonian
were tuned to the equation of state of infinite nuclear matter and
to properties of finite nuclei (such as their binding energy and
their root mean square radius). If, however, the energy of the
nucleons within the nucleus is minimized according to the Hamiltonian
in a self-consistent fashion, then the nucleus would collapse to
a single point in momentum space because the Pauli-principle
has not been taken into account in the Hamiltonian.
One can use a so-called Pauli-Potential \cite{wilets77a} in the Hamiltonian.
Its advantage is that the initialized nuclei remain absolutely stable
whereas in the conventional initialization and propagation without
the Pauli-Potential the nuclei start evaporating single nucleons after
approximately 20 - 30 fm/c. A drawback of the potential is that the
kinetic momenta of the nucleons are not anymore equivalent to their
canonic momenta, i.e. the nucleons carry the correct Fermi-momentum,
but their velocity is zero. Furthermore, the Pauli-Potential leads to
a wrong specific heat and changes the dynamics of fragment formation.

Impact parameter of a collision is sampled according to the quadratic
measure ($dW\sim~bdb$). At given impact parameter centers of projectile
and target are placed along the collision axis in such a manner that
a distance between surfaces of the projectile and the target is equal
to 3 $fm$. Momenta of nucleons are transformed in the system where
the projectile and target have equal velocities directed in different
directions of the axis. After that the time propagation starts.
During the calculation each particle is checked at the beginning of each
time step whether it will collide within that time step.
A collision between two hadrons will occur if
$d<\sqrt{\sigma_{\rm tot}/\pi}$, where $d$ and $\sigma_{\rm tot}$
are the impact parameter of the hadrons and the total cross-section of the
two hadrons, respectively. After each binary collision or decay the
outgoing particles are checked for further collisions within the respective
time step.

In the UrQMD model the total cross-section $\sigma_{\rm tot}$ depends
on the isospins of colliding particles, their flavor and the c.m. energy.
The total and elastic proton-proton and proton-neutron cross sections
are well known \cite{PDG96}. Since their functional dependence on
$\sqrt{s}$ shows at low energies a complicated shape, UrQMD uses a
table-lookup for those cross sections. The neutron-neutron cross section
is treated as equal to the proton-proton cross section (isospin-symmetry).
In the high energy limit ($\sqrt{s} \ge 5$~GeV) the CERN/HERA
parameterization for the proton-proton cross section is used
\cite{PDG96}.

Baryon resonances are produced in two different ways, namely
\begin{itemize}
\item[\bf i)] {\it hard production\/}:
N+N$\rightarrow \Delta$N,\ $\Delta\Delta$,\ N$^*$N, etc.
\item[\bf ii)] {\it soft production\/}: $\pi^-+$p$\rightarrow
\Delta^0$,\ K$^-$+p$\rightarrow \Lambda^*$...
\end{itemize}
The formation of $s$-channel resonances is fitted to measured data.
Partial cross-sections are used to calculate the relative weights for
the different channels.

There are six channels of the excitation of
non-strange resonances in the UrQMD model, namely $NN \rightarrow N
\Delta_{1232}, N N^{\ast}, N \Delta^{\ast}, \Delta_{1232}
\Delta_{1232}, \Delta_{1232} N^{\ast}$, and $\Delta_{1232}
\Delta^{\ast}$.
The $\Delta_{1232}$ is explicitly listed,
whereas higher excitations of the $\Delta$ resonance
have been denoted as $\Delta^{\ast}$.
For each of these 6 channels specific assumptions have been made
with respect to the form of the matrix element, and the free
parameters have been adjusted to the available experimental data.

Meson-baryon cross-sections are dominated by the formation of $s$-channel
resonances, i.e. the formation of a transient state of mass
$m=\sqrt{s_{hh}}$, containing the total c.m. energy of the two incoming
hadrons. On the quark level such a process implies that a quark from
the baryon annihilates an antiquark from the incoming meson. Below 2.2 GeV
c.m. energy intermediate resonance states get excited. At higher energies
the quark-antiquark annihilation processes become less important. There,
$t$-channel excitations of the hadrons dominate, where the exchange of
mesons and Pomeron exchange determines the total cross-section of the
MB interaction \cite{donna}.

To describe the total meson-meson reaction cross-sections,
the additive quark model and the principle of detailed
balance, which assumes the reversibility of the particle interactions
are used.

Resonance formation cross sections from the
measured decay properties of the possible resonances up to
c.m. energies of 2.25~GeV/$c^2$ for baryon resonance and 1.7~GeV/$c^2$
in the case of MM and MB reactions have been calculated based on the
principle. Above these energies collisions are modeled by the formation
of $s$-channel string or, at higher energies (beginning at $\sqrt s =3$~GeV),
by one/two $t$-channel strings. In the strangeness channel
elastic collisions are possible for those meson-baryon combinations
which are not able to form a resonance, while the creation of
$t$-channel strings is always possible at sufficiently large
energies. At high collision energies both cross section become equal
due to quark counting rules.

A parameterization proposed by Koch and Dover \cite{kochP89a}
is used in UrQMD model for baryon-antibaryon annihilation cross section.
It is assumed that the antiproton-neutron annihilation cross section is
identically to the antiproton-proton annihilation cross section.

The total and elastic proton-antiproton cross sections are treated
according to the CERN/HERA parameterization:
\begin{equation}
\sigma(p)\,=\,{\rm A} + {\rm B}\, p^n + {\rm C}\,{\rm ln}^2(p)
        + {\rm D}\,{\rm ln}(p)\, ,
\end{equation}
with the laboratory-momentum $p$ in GeV/c.
The parameters are listed in \cite{UrQMD2}.

For momenta $p_{lab}<5$~GeV/c, UrQMD uses another parameterization
to obtain a good fit to the data:
\begin{equation}
\sigma_{\rm tot}(p) =\left\{ \begin{array}{ll}
75.0+43.1 p^{-1} + 2.6 p^{-2} -3.9 p &\quad :\quad 0.3<p<5\\
271.6\exp (-1.1\, p^2)  & \quad:\quad p< 0.3
\end{array} \right.
\end{equation}

\begin{equation}
\sigma_{\rm el}=\left\{ \begin{array}{ll}
31.6+18.3 p^{-1} -1.1 p^{-2} - 3.8 p &\quad :\quad 0.3<p< 5\\
78.6 &\quad :\quad p< 0.3\;
\end{array} \right.
\end{equation}

For low lab-momenta the annihilation cross section is dominant.
The sum of annihilation and elastic cross section, however, is
smaller than the total cross section:
\begin{equation}
 \Delta\sigma = \sigma_{\rm tot}-\sigma_{\rm el}-\sigma_{\rm ann}
\end{equation}
This difference is called ``diffractive'' cross section in UrQMD,
$\sigma_{\rm diff}=\Delta \sigma$, and is used to excite one
(or both) of the collision partners to a resonance or to
a string. In the string case the same excitation scheme as for
proton-proton reactions is used. For high energies  the ``diffractive''
cross section is the dominant contribution to the total
antiproton-proton cross section.

The final state of a baryon-antibaryon annihilation is generated
via the formation of two meson-strings. The available c.m. energy
of the reaction is distributed in equal parts to the two strings
which decay in the rest frame of the reaction.
On the quark level this procedure implies the annihilation of a
quark-antiquark pair and the reordering of the remaining
constituent quarks into newly produced hadrons (additionally taking
sea-quarks into account). This model for the baryon-antibaryon
annihilation thus follows the topology of a {\em rearrangement}-graph.

The hadron-hadron interactions at high energies are simulated in 3 stages.
According to the cross sections the type of interaction is  defined:
elastic, inelastic, antibaryon-baryon annihilation etc. In the case of
inelastic collision with string excitation the kinematical characteristics
of strings are determined. The strings between quark and diquark
(antiquark) from the same hadron are produced. The strings have the continuous
mass distribution $f(M) \propto 1/M^2$ with the masses $M$, limited by
the total collision energy $\sqrt{s}$: $M_1+M_2 \le \sqrt{s}$.
The rest of the $\sqrt{s}$ is equally distributed
between the longitudinal momenta of two produced strings.

The second stage of h-h interactions is connected with
string fragmentation. The string break-up is treated
iteratively: String $\rightarrow$ hadron + smaller string.
A quark-antiquark (or a diquark-antidiquark) pair is created and
placed between leading constituent quark-antiquark (or diquark-quark)
pair.  Then a hadron is formed randomly on one of the end-points of the
string. The quark content of the hadron determines its species
and charge. In case of resonances the mass is determined according to
a Breit-Wigner distribution. Finally, the energy-fraction of the string
which is assigned to the newly created hadron is determined:
After the hadron has been stochastically assigned a transverse momentum,
the fraction of longitudinal momentum transferred from the string
to the hadron is determined by the fragmentation function.
The conservation laws are fulfilled. The diquark is permitted to
convert into mesons via the breaking of the diquark link.

This iterative fragmentation process is repeated until the remaining
energy of the string gets too small for a further fragmentation.

The fragmentation function $f(x,m_t)$ represents the probability distribution
for hadrons with the transverse mass $m_t$ to acquire the longitudinal
momentum fraction $x$ from the fragmenting string. One of the most
common fragmentation functions is the one used in the LUND model
\cite{andersson83a}.
In UrQMD, different fragmentation functions are used for leading nucleons
and newly produced particles, respectively:
\begin{eqnarray}
\label{fuqmd1}
f(x)_{\rm nuc} &=& \exp\left(-\frac{(x-B)^2}{2\,A^2}\right)
                        \quad \mbox{for leading nucleons} \\
\label{fuqmd2}
f(x)_{\rm prod} &=& (1-x)^2
                        \quad \mbox{for produced particles}
\end{eqnarray}
with $A=0.275$ and $B=0.42$.
The fragmentation function $f(x)_{\rm prod}$, used for newly produced
particles, is the well-known Field-Feynman
fragmentation function \cite{field77a,field78a}.

The fragmentation scheme determines formation time of created hadrons. Though
there are various possibility (for details see Ref.
\cite{UrQMD1,UrQMD2}).

After the fragmentation, decay of the resonances proceeds according to the
branching ratios compiled by the Particle Data Group \cite{PDG96}.
The resonance decay products have isotropical distributions in the rest
frame of the resonance. If a resonance is among the outgoing particles,
its mass must first be determined  according to a Breit-Wigner
mass-distribution. If the resonance decays into $N > 2$ particles,
then the corresponding $N-$body phase space is used to calculate their $N$
momenta stochastically.

The Pauli principle is applied to hadronic collisions or decays by
blocking the final state if the outgoing phase space is occupied.

The UrQMD collision term contains 55 different baryon species
(including nucleon, delta and hyperon resonances with masses up
to 2.25 GeV/$c^2$) and 32 different meson species (including strange
meson resonances), which are supplemented by their corresponding
anti-particle and all isospin-projected states. The states can either
be produced in string decays, s-channel collisions or resonance decays.
For excitations with higher masses than 2 GeV/$c^2$ a string picture is
used. Full baryon/antibaryon symmetry is included:
The number of implemented baryons therefore defines the number
of antibaryons in the model and the antibaryon-antibaryon interaction
is defined via the baryon-baryon interaction cross sections.

Elementary cross sections are fitted to available proton-proton or
pion-proton data. Isospin symmetry is used when possible in order
to reduce the number of individual cross sections which have to
be parameterized or tabulated.

The UrQMD model reproduces nicely various properties of hadron-hadron
and nucleus-nucleus interactions.

\section*{Appendix 2: List of used papers}
\begin{enumerate}

\item "Microscopic Models for  Ultrarelativistic Heavy Ion Collisions",\\
S.A. Bass et al., {\it Prog. Part. Nucl. Phys.},{\bf 41} (1998) 225;
nucl-th/9803035.

\item "J/$\Psi$ Suppresion in Heavy Ion Collisions - Interplay of Hard and Soft QCD Processes",\\
C. Spieles, R. Vogt, L. Gerland, S.A. Bass, M. Bleicher, L. Frankfurt, M. Strikman,
W. Greiner, {\it LBL-42410}, 1998;
hep-ph/9810486.

\item "Relativistic Hadron Hadron Collisions in the Ultrarelativistic Quantum Molecular
Dynamics Model",\\
M. Bleicher et al., {\it J. Phys.}, {\bf G25} (1999) 1859;
hep-ph/9909407.

\item "Local Equilibrium in Heavy Ion Collisions: Microscopic Model Versus Statistical Model
Analysis",\\
L.V. Bravina et al., {\it Phys. Rev.} {\bf C60} (1999) 024904;
hep-ph/9906548.

\item "Modeling J/$\Psi$ Production and Absorption in a Microscopic Nonequilibrium Approach",\\
C. Spieles, R. Vogt, L. Gerland, S.A. Bass, M. Bleicher, Horst Stoecker, W. Greiner,
{\it Phys. Rev.} {\bf C60} (1999) 054901;
hep-ph/9902337.

\item "Physics Opportunities at RHIC and LHC",\\
S. Scherer et al., {\it Proceedings of the Klaus Kinder-Geiger Day.}, 1999;
hep-ph/9903392.

\item "Directed and Elliptic Flow",\\
Sven Soff, Steffen A. Bass, Marcus Bleicher, Horst Stoecker, Walter Greiner,
nucl-th/9903061.

\item "Direct Emission of Multiple Strange Baryons in Ultrarelativistis Heavy Ion Collisions
from the Phase Boundary",\\
A. Dumitru, S.A. Bass, M. Bleicher, Horst Stoecker, W. Greiner,
{\it Phys. Lett.}  {\bf B460} (1999) 411;
nucl-th/9901046.

\item "Sideward Flow in AU + AU Collisions Between 2-A-GeV and 8-A-GeV",\\
By E895 Collaboration (H. Liu et al.),
{\it Phys. Rev. Lett.} {\bf 84} (2000) 5488;
nucl-ex/0005005.

\item "Enhanced Anti-Proton Production in PB(160-GeV/A) + PB Reactions: Evidence for Quark
Gluon Matter?",\\
M. Bleicher, M. Belkacem, S.A. Bass, S. Soff, Horst Stoecker,
{\it Phys. Lett.} {\bf B485} (2000) 133;
hep-ph/0004045.

\item "Baryonic Contributions to the Dilepton Spectra in Relativistic Heavy Ion Collisions",\\
M. Bleicher, A.K. Dutt-mazumder, C. Gale, C.M. Ko, V. Koch, {\it LBNL-45199}, 2000;
nucl-th/0004044.

\item "Direct Photon Production in 158-A-GeV PB-208 + PB-208 Collisions",\\
By WA98 Collaboration (M.M. Aggarwal et al.),
nucl-ex/0006007.

\item "Nuclear Transparency in Heavy Ion Collisions at 14.6-GeV/Nucleon",\\
H.M. Ding, P. Glassel, J. Hufner,
{\it Nucl. Phys.} {\bf A692} (2001) 549;
nucl-th/0008002.

\item "Enhanced Strange Particle Yields - Signal of a Phase of Massless Particles?",\\
Sven Soff, D. Zschiesche, M. Bleicher, C. Hartnack, M. Belkacem, L. Bravina, E. Zabrodin,
S.A. Bass, Horst Stoecker, W. Greiner,
{\it J. Phys.} {\bf G27} (2001) 449;
nucl-th/0010103.

\item "Strangennes in Dense Nuclear Matter: a Review of AGS Results",\\
Fuqiang Wang,
{\it J. Phys.} {\bf G27} (2001) 283;
nucl-ex/0010002.

\item "Current Status of Quark Gluon Plasma Signals",\\
D. Zschiesche et al.,
{\it Heavy Ion Phys.} {\bf 14} (2001)425;
nucl-th/0101047.

\item Rapidity Dependence of Anti-Proton to Proton Ratios in AU+A Collisions
at S(NN)**(1/2) = 130-GEV"\\
BRAHMS Collaboration (I.G. Bearden et al.),
{\it Phys. Rev. Lett.} {\bf 87} (2001) 112305;
nucl-ex/0106011.

\item "Charged Particle Densities from AU+AU Collisions at S(NN)**(1/2) = 130-GEV",\\
BRAHMS Collaborations (I.G. Bearden et al.),
{\it Phys. Lett.} {\bf B523} (2001) 227;
nucl-ex/0108016.

\item "Strangeness Production in Microscopic Transport Models",\\
Steffen A. Bass,
{\it J. Phys.} {\bf G28} (2002) 1543;
nucl-th/0112046.

\item "Lambda Production in 40-A-GEV/C PB AU Collisions",\\
CERES Collaboration (Wolfgang Schmitz et al.),
{\it J. Phys.} {\bf G28} (2002) 1861;
nucl-ex/0201002.

\item "Dileptons and Photons from Coarse Grained Microscopic Dynamics and Hydrodynamics
Compared to Experimental Data",\\
P. Huovinen, M. Belkacem, P.J. Ellis, Joseph I. Kapusta,
{\it Phys. Rev.} {\bf C66} (2002) 014903;
nucl-th/0203023.

\item "Probing Hadronization with Strangeness",\\
S.A. Bass, M. Bleicher, J. Aichelin, F. Becattini, A. Keranen, F.M. Liu, K. Redlich,
K. Werner,
nucl-th/0204049

\item "Hardonic Observables from SIS to SPS Energies: Anything Strange with Strangeness?",\\
H. Weber, E.L. Bratkovskaya, W. Cassing, Horst Stoecker,
{\it Phys. Rev.} {\bf C67} (2003) 014904;
nucl-th/0209079.

\item "Baryon Stopping as a Probe for Highly Excited Nuclear Matter",\\
H. Weber, E.L. Bratkovskaya, Horst Stoecker,
nucl-th/0205032.

\item "Exploring Isospin, Strangeness and Charm Distillation in Nuclear Collisions",\\
M. Reiter, E.L. Bratkovskaya, M. Bleicher, W. Cassing, Horst Stoecker,
{\it Nucl. Phys.} {\bf A722} (2003) 142;
nucl-th/0301067.

\item "Search for Deconfinement in NA49 at the CERN SPS",\\
Peter Seyboth et al.,
{\it Heavy Ion Phys.} {\bf 15} (2002) 257; {\it Pramana} {\bf 60} (2003) 725;
hep-ex/0206046.

\item "Event by Event Fluctuations of the Mean Transverse Momentum in 40, 80 AND 158 A
GEV/C PB-AU Collisions",\\
CERES Collaboration (D. Adamova et al.),
{\it Nucl. Phys.} {\bf A727} (2003) 97;
nucl-ex/0305002.

\item "Simulation of Anti-Proton Nucleus Interactions in the Framework of the URQMD Model",\\
A.S. Galoyan, A. Polanski,. JINR-E1-2003-125;
hep-ph/0304196.

\item "Model Dependence of Lateral Distribution Functions of High Energy Cosmic Ray Air Shower",\\
Hans-Joachim Drescher, Marcus Bleicher, Sven Soff, Horst Stoecker,
{\it Astropart. Phys.} {\bf 21} (2004) 87;
astro-ph/0307453.

\item "Evidence for Nonhadronic Degrees of Freedom in the Transverse Mass Spectra of Kaons
from Relativistic Nucleus Nucleus Collisions?,\\
E.L. Bratkovskaya, S. Soff, Horst Stoecker, M. van Leeuwen, W. Cassing,
{\it Phys. Rev. Lett.} {\bf 92} (2004) 032302;
nucl-th/0307098.

\item :Universal Transition Curve in Pseudorapidity Distribution",\\
Sangyong Jeon, Vasile Topor Pop, Marcus Bleicher,
{\it Phys. Rev.} {\bf C69} (2004) 044904;
nucl-th/0309077.

\item "Multistrange Baryon Production in AU+AU Collisions Near Threshold",\\
Gebhard Zeeb, Manuel Reiter, Marcus Bleicher,
{\it Phys. Lett.} {\bf B586} (2004) 297;
nucl-th/0312015.

\item "Strangeness Dynamics in Relativistic Nucleus Nucleus Collision",
E.L.~Bratkovskaya, M.~Bleicher, W.~Cassing, M.~van~Leeuwen, M.~Reiter, S.~Soff,
Horst~Stoecker, H.~Weber,
{\it Prog. Part. Nucl. Phys.} {\bf 53} (2004) 225;
nucl-th/0312048

\item "Strangeness Dynamics and Transverse Pressure in Relativistic Nucleus-Nucleus Collisions",
E.L.~Bratkovskaya, M.~Bleicher, M.~Reiter, S.~Soff, Horst~Stoecker, M.~van~Leeuwen,
S.A.~Bass, W.~Cassing,
{\it Phys. Rev.} {\bf C69} (2004) 054907;
nucl-th/0402026.

\item "Omega- and Anti-Omega+ Production in Central PB+PB Collisions at 40-AGeV AND 158-AGeV",\\
NA49 Collaboration (C. Alt et al.),
{\it Phys. Rev. Lett.} {\bf 94} (2005) 192301;
nucl-ex/0409004.

\item "Low Energy Hadronic Interaction Models",\\
Dieter Heck,
{\it Nucl. Phys. Proc. Suppl.} {\bf 151} (2006) 127;
astro-ph/0410735.

\item "Transverse Prssure and Strangeness Dynamics in Relativistic Heavy Ion Reactions",\\
M. Bleicher, E. Bratkovskaya, S. Vogel, X. Zhu,
{\it J. Phys.} {\bf G31} (2005) S709;
hep-ph/0503252.

\item "Is the Existence of a Softest Point in the Directed Flow Excitation Function an
Unambiguous Signal for the Creation of a Quark Gluon Plasma?",\\
Marcus Bleicher, Jorg Aichelin,
{\it Phys. Lett.} {\bf B612} (2005) 201;
nucl-th/0205069.

\item "Reconstructing  Rho0 and Omega Mesons from Non-Leptonic Decays in C+C at 2-A-GeV",\\
Sascha Vogel, Marcus Bleicher,
nucl-th/0509105.

\item "Excitation Function of the Longitudinal Expansion in Central Nuclear Collisions",\\
Marcus Bleicher,
hep-ph/0504207.

\item "Summary of Theoretical Contributions",\\
Horst Stoecker,
nucl-th/0506013.

\item "NA49 Results on Hadron Production: Indications of the Onset of Deconfinement?",\\
Benjamin Lungwitz,
nucl-ex/0509041.

\item "Directed Flow in AU+AU Collisions at S(NN)**(1/2) = 62 GeV",\\
STAR Collaboration (J. Adams et al.),
nucl-ex/0510053.

\item "Elliptic Flow Analysis at RHIC: Fluctuations vs. Non-Flow Effects",\\
Xiang-lei Zhu, Marcus Bleicher, Horst Stocker,
nucl-th/0509081.

\item "Particle Number Fluctuations in High Energy Nucleus-Nucleus Collisions from
Microscopic Transport Approaches",\\
V.P. Konchakovski, S. Haussler, M.I. Gorenstein, E.L. Bratkovskaya, M. Bleicher, H. Stocker,
nucl-th/0511083.

\item "Energy and Centrality Dependence of Antiproton and Proton Production in Relativistic
PB+PB Collisions at the CERN SPS",\\
NA49 Collaboration,
nucl-ex/0512033.

\end{enumerate}

\end{document}